\documentclass[a4paper,12pt]{article}
\usepackage{graphicx,amssymb,bm,latexsym,color,cite,epsf}
\usepackage{amsmath,lmodern}
\usepackage{appendix}
\usepackage{comment}
\usepackage{xcolor}
\usepackage[normalem]{ulem}
\usepackage{hyperref}
\usepackage{mathrsfs}
\usepackage[utf8]{inputenc}
\usepackage{ulem}
\usepackage{xcolor}
\usepackage[nameinlink]{cleveref}
\DeclareRobustCommand{\erase}{\bgroup\markoverwith{\textcolor{red}{\rule[.5ex]{2pt}{0.4pt}}}\ULon}

\hypersetup{
colorlinks=true,
citecolor=blue,
citebordercolor=red,
linktoc=all,
linkcolor=blue,
urlcolor=blue
}

\makeatletter
\@addtoreset{equation}{section}

\makeatother
\newcommand{\bea}{\begin{eqnarray}}
\newcommand{\eea}{\end{eqnarray}}
\newcommand{\vs}[1]{\vspace{#1 mm}}
\newcommand{\hs}[1]{\hspace{#1 mm}}
\renewcommand{\a}{\alpha}
\renewcommand{\b}{\beta}
\newcommand{\G}{\Gamma}
\renewcommand{\d}{\delta}
\newcommand{\e}{\epsilon}
\newcommand{\s}{\sigma}

\newcommand{\vp}{\varphi}
\newcommand{\la}{\lambda}
\newcommand{\pa}{\partial}

\newcommand{\nn}{\nonumber\\}
\newcommand{\p}{\partial}

\newcommand{\br}{\bar R}
\newcommand{\bg}{\bar g}

\newcommand{\bp}{\bar \phi}

\newcommand{\bnabla}{\bar\nabla}
\newcommand{\tr}{{\rm tr}}
\newcommand{\Tr}{{\rm Tr}}

\newcommand{\pmat}[1]{\begin{pmatrix}#1\end{pmatrix}}
\newcommand{\df}{{\rm d}}
\allowdisplaybreaks

\begin{document}
\begin{flushright}
NITEP 254 \\
\today
\end{flushright}

\renewcommand{\thefootnote}{\fnsymbol{footnote}}
\newbox{\ORCIDicon}
\sbox{\ORCIDicon}{\large  \includegraphics[width=0.8em]{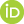}}

\begin{center}
{\Large\bf
Essential Renormalization Group Equation \\ \vs{3}
for Gravity coupled to a Scalar field
}
\vs{5}

{\large
Nobuyoshi Ohta\,\href{https://orcid.org/0000-0002-8265-8636}{\usebox{\ORCIDicon}},$^{1,}$\footnote{e-mail address: ohtan.gm@gmail.com}
and Masatoshi Yamada\,\href{https://orcid.org/0000-0002-1013-8631}{\usebox{\ORCIDicon}}$^{2,}$\footnote{e-mail address: m.yamada@kwansei.ac.jp}
} \\
\vs{5}

${}^1${\em Nambu Yoichiro Institute of Theoretical and Experimental Physics (NITEP), \\
Osaka Metropolitan University, Osaka 558-5585, Japan}
\vs{3}

${}^2${\em Department of Physics and Astronomy, Kwansei Gakuin University,
Sanda, Hyogo, 669-1330, Japan}

\vs{5}
{\bf Abstract}
\end{center}

We study the essential renormalization group equation, in which inessential couplings are removed via field redefinitions, for Einstein gravity coupled to a massive scalar field in the presence of a cosmological constant.
Our results indicate that perturbatively nonrenormalizable terms can be eliminated due to the cosmological term, in contrast to the case of perturbation around flat spacetime.
We find a nontrivial fixed point for the Newton coupling and the cosmological term.

\renewcommand{\thefootnote}{\arabic{footnote}}
\setcounter{footnote}{0}

\section{Introduction and motivation}

One of the long-standing challenges in theoretical physics is the formulation of a consistent quantum theory of gravity.
Quantum Einstein gravity is known to be nonrenormalizable within perturbation theory. 
In the seminal work~\cite{tHooft:1974toh}, one-loop perturbative calculations are made with the Einstein-Hilbert action, and the result revealed that quantum fluctuations of the metric generate divergent counterterms involving the Ricci scalar squared $R^2$ and the Ricci tensor squared $R_{\mu\nu}R^{\mu\nu}$. Since these terms are absent in the bare Einstein-Hilbert action, this result naively suggests the nonrenormalizability of Einstein gravity.
However, these divergent terms are proportional to the classical equations of motion for the background metric and can be eliminated on-shell. 
More precisely, they can be removed via a field redefinition~\cite{Hawking:1979ig}, indicating that they do not affect physical observables such as the $S$-matrix~\cite{Chisholm:1961tha,Kamefuchi:1961sb}.
Such counterterms are referred to as \textit{redundant} operators, and their coefficients are known as \textit{inessential} couplings.
This fact implies that quantum Einstein gravity is on-shell renormalizable at the one-loop level.

It is also found in Ref.~\cite{tHooft:1974toh} that when a massless free scalar field is (minimally) coupled to gravity without the cosmological constant, the theory is no longer on-shell renormalizable even at the one-loop level, as some of the resulting counterterms cannot be removed using the classical field equations.
Furthermore, at the two-loop level, a cubic Riemann tensor term arises as a counterterm, the so-called Goroff-Sagnotti term, which cannot be eliminated by the field redefinition~\cite{Goroff:1985sz,Goroff:1985th,vandeVen:1991gw}.
This is the definite evidence of the perturbative nonrenormalizability of the Einstein theory.

On the other hand, the asymptotic safety scenario for quantum gravity in four-dimensional spacetime \cite{Reuter:1996cp,Souma:1999at} has recently attracted much attention.
This approach employs the functional renormalization group (FRG) to search for nontrivial ultraviolet (UV) fixed points of couplings in the theory.
(The first attempt to calculate the beta functions perturbatively for gravity with higher curvatures was made in Refs.~\cite{Julve:1978xn,Fradkin:1981iu,Avramidi:1985ki} and found only the Gaussian fixed point in the couplings of quadratic curvature terms.)
If a quantum field theory can be consistently restricted to a finite set of operators with well-defined UV fixed points under the renormalization group (RG) flow, the theory becomes scale invariant at high energies, and the UV divergences may be brought under control. 
To date, substantial evidence has been accumulated in support of this scenario, both for pure gravity~\cite{
Reuter:2001ag,Lauscher:2002sq,deBerredoPeixoto:2004if,Ohta:2013uca,Ohta:2015zwa,Codello:2006in,Machado:2007ea,Codello:2007bd,Benedetti:2012dx,Falls:2013bv,Benedetti:2013jk,Falls:2014tra,Falls:2017lst,Falls:2018ylp,DeBrito:2018hur,Ohta:2015efa,Ohta:2015fcu,Gies:2016con,Falls:2016msz,Christiansen:2016sjn,Denz:2016qks,Falls:2020qhj,Sen:2021ffc,Kawai:2023rgy,Kawai:2024rau,Benedetti:2009rx,Benedetti:2009gn} and for gravity-matter systems~\cite{Percacci:2003jz,Narain:2009fy,Narain:2009gb,Eichhorn:2012va,Henz:2013oxa,Henz:2016aoh,Wetterich:2019rsn,Dona:2013qba,Labus:2015ska,Oda:2015sma,Hamada:2017rvn,Eichhorn:2017als,Pawlowski:2018ixd,Wetterich:2019zdo,Meibohm:2015twa,Christiansen:2017cxa,Pastor-Gutierrez:2022nki,Christiansen:2017bsy,Ohta:2021bkc}.
See also reviews~\cite{Niedermaier:2006wt,Niedermaier:2006ns,Percacci:2007sz,Reuter:2012id,Codello:2008vh,Eichhorn:2017egq,Percacci:2017fkn,Eichhorn:2018yfc,Reuter:2019byg,Bonanno:2020bil,Reichert:2020mja,Eichhorn:2022jqj,Eichhorn:2022gku,Pawlowski:2020qer,Pawlowski:2023gym}.

In the standard FRG approach, we should include all possible effective operators---including redundant operators---in the effective action and study which operators are relevant and which are irrelevant. However, in practice, it is impossible to deal with all effective operators. What we can do is start the analysis of the system with lower-dimensional operators and then include more operators to determine if the additional operators are relevant or not, and expect that this process ends at a certain level.

Recently, an alternative formulation was proposed that focuses on essential couplings within the FRG framework~\cite{Baldazzi:2021ydj}. 
In this approach, redundant operators are systematically eliminated through the field redefinitions, and the flow equations are derived exclusively for the essential couplings. 
This method, called the essential RG equation, not only reduces computational complexity but also offers a clearer perspective on the fixed-point structure of the system.

In this context, Ref.~\cite{Baldazzi:2021orb} suggests that the quadratic curvature terms, namely $R^2$ and $R_{\mu\nu}R^{\mu\nu}$, in asymptotically safe gravity may be redundant operators, removable through suitable field redefinitions. In other words, the counterterms that appear at one-loop order may not correspond to essential couplings. This suggestion aligns with the result of Ref.~\cite{tHooft:1974toh}, where such terms were shown to vanish by the use of field equations as mentioned before. As a consequence, only the Einstein-Hilbert term and the cosmological constant remain as essential and relevant operators in the asymptotic safety framework. Building on the essential RG equation approach, Ref.~\cite{Baldazzi:2023pep} investigated the Goroff-Sagnotti term, nonremovable by the field redefinitions~\cite{tHooft:1974toh}, and showed that the term is an irrelevant operator.

It still remains to be seen if this approach is successful in the gravity coupled to matter fields.
The purpose of the present paper is to extend this approach to the system of gravity coupled to a scalar field as a first step.\footnote{
An attempt motivated by this idea was made in Ref.~\cite{Knorr:2022ilz} which studied gravity coupled to a shift-invariant scalar field.
}
This issue gets its urgency in view of the fact that such a scalar-gravity system is known to be nonrenormalizable in perturbation theory even at one-loop level~\cite{tHooft:1974toh}. 
In particular, we intend to explore that the problem may be overcome due to the existence of the cosmological constant term. 
A similar use of the cosmological term to eliminate the counterterms was made in Ref.~\cite{Solodukhin:2015ypa}.
Extension to gravity with more scalars or spin-1/2 fermions and spin-1 gauge bosons should be~straightforward.

This paper is organized as follows. In \Cref{pt}, we summarize the work on the counterterms in perturbation theory of Einstein theory without and with scalar fields. In \Cref{'tHooft-Veltman}, we discuss the calculations in the Einstein theory and show that the theory is renormalizable on-shell at one-loop level. This can also be understood as the renormalizability by the field redefinition. We also discuss that the theory is not renormalizable at two-loop level, and is already so at one-loop level if the gravity is coupled to scalar fields. In \Cref{higher}, we briefly discuss the properties of gravity theory with quadratic curvature terms.
In \Cref{ass}, we explain the asymptotic safety scenario.
In \Cref{FRGST}, we start our study of the Einstein gravity with vacuum energy coupled to a scalar field in the usual FRG framework. In \Cref{scalar-tensor}, we set up the system we consider in this paper. We then derive the Hessians for the system in \Cref{hessian} and find the results for FRG equations in \Cref{standard}. In \Cref{essential}, we introduce the formulation of essential RG, and show that the nonrenormalizable counterterms may be removed by the field redefinition.
Finally \Cref{summary} is devoted to the summary of our results.

Some computational details are relegated to the appendixes.
In \Cref{appsec:Variations}, we summarize the Hessians for the gravity sector.
In \Cref{appsec:Heat kernel expansion}, we collect the formulas for the heat kernel expansion.
Finally, in \Cref{eom}, we give the equations of motion for our system.

\section{Renormalizability in pure gravity and scalar-tensor theory}
\label{pt}

In this section, we discuss renormalizability in quantum gravity and scalar-tensor theories within the framework of perturbative quantum field theory. We begin by reviewing perturbative calculations in pure Einstein gravity and in Einstein gravity coupled to a free scalar field. In both cases, we consider the formulation without assuming the presence of a cosmological constant.
Second, we comment on higher derivative gravity in which the bare action contains the quadratic terms $R^2$ and $R_{\mu\nu}R^{\mu\nu}$ in addition to the Einstein-Hilbert term.
Finally, we explain the idea of asymptotically safe gravity and summarize its current status. 

\subsection{Perturbative calculations}
\label{'tHooft-Veltman}

In this subsection, we review the work of 't Hooft and Veltmann~\cite{tHooft:1974toh} who have discussed the perturbative renormalizability of quantum-gravity models.
To this end, let us first discuss the pure gravity based on the Einstein-Hilbert action in Euclidean spacetimes,
\begin{align}
S_\text{EH} = -\int \df^dx\sqrt{g}\, R\,,
\label{eq:EH action}
\end{align}
where we have set $16\pi G_N=1$ for simplicity only in this section.
The perturbative calculation has been performed around a fixed background spacetime $\bar g_{\mu\nu}$, i.e., we split the metric as $g_{\mu\nu}=\bar g_{\mu\nu}+h_{\mu\nu}$ where $h_{\mu\nu}$ stands for the fluctuating metric field.
Here the gauge-fixing term and the corresponding ghost term are given respectively by
\begin{align}
S_{\rm gf} &= -\frac{1}{2}\int \df^dx\sqrt{\bar g} \left( \bar\nabla_\nu h_{\mu}{}^\nu-\frac{1}{2} \bar\nabla_{\mu} h \right)\left( \bar\nabla_\rho h^{\mu\rho} -\frac{1}{2} \bar\nabla^{\mu} h \right),\\
S_{\rm gh} &=- \int \df^dx\sqrt{\bar g}\,  b^\mu\left[ -\bar\square + \bar R_{\mu\nu}  \right]c^\nu,
\end{align}
where $\bar\square=\bar\nabla_\mu \bar\nabla^\mu$ and $c^\mu$ and $b_\mu$ are the Faddeev-Popov (FP) ghost and antighost, respectively.
Here and in what follows, a bar indicates that the quantity is evaluated on the background; the
indices are raised, lowered, and contracted by the background metric $\bar g$, the covariant derivative $\bar \nabla$
is constructed with the background metric.

It has been shown in Ref.~\cite{tHooft:1974toh} that the perturbative one-loop calculation entails the following counter Lagrangian with higher dimensional operators:
\begin{align}
\Delta \mathcal L\Big|^\text{1-loop}_\text{EH}
=\frac{\sqrt{\bar g}}{8\pi^2 \varepsilon}\left( \frac{1}{120}\bar R^2 + \frac{7}{20}\bar R_{\mu\nu}\bar R^{\mu\nu} \right),
\end{align}
where the dimensional regularization with $\varepsilon=d-4$ has been used.
Here, a key point is that the Riemann tensor squared $\bar R_{\mu\nu\rho\sigma}\bar R^{\mu\nu\rho\sigma}$ has been eliminated by using the topological nature of the Gauss-Bonnet term in four-dimensional spacetimes,
\begin{align}
\mathfrak{G}\equiv \bar R_{\mu\nu\rho\sigma}\bar R^{\mu\nu\rho\sigma} - 4\bar R_{\mu\nu}\bar R^{\mu\nu} + \bar R^2= \text{(total derivative term)} ,
\label{eq:Gauss-Bonnet term}
\end{align}
which vanishes within the spacetime integral $\int\df^4x\sqrt{\bar g}$.
Because the Einstein-Hilbert action~\labelcref{eq:EH action} does not contain the higher-dimensional operators $\bar R^2$ and $\bar R_{\mu\nu}\bar R^{\mu\nu}$, one might think that the Einstein-Hilbert theory is not perturbatively renormalizable.
However, we invoke the equation of motion (i.e., the Einstein equation) for the background metric field
\begin{align}
0=\frac{1}{\sqrt{\bar g}}\frac{\delta S_\text{EH}}{\delta \bar g^{\mu\nu}} = \bar R_{\mu\nu} -\frac12 \bar g_{\mu\nu} \bar R\,,
\end{align}
for which taking the trace, we have $\bar R=0$ and $\bar R_{\mu\nu}=0$.
This fact leads to $\Delta \mathcal L|^\text{1-loop}_\text{EH}=0$ on-shell and therefore we conclude that the Einstein-Hilbert theory is perturbatively renormalizable at the one-loop level.

This property is often referred to as ``on-shell'' renormalizability. More precise formulation is by the field redefinition~\cite{Chisholm:1961tha,Kamefuchi:1961sb,
Hawking:1979ig}. When we change any unrenormalized coupling parameter $\gamma_0$ by an infinitesimal amount $\e$, the whole Lagrangian changes by
\bea
{\cal L} \to {\cal L}+\e \frac{\pa \cal L}{\pa \gamma_0}.
\eea
Suppose we try to reproduce this change by a mere redefinition of fields
\bea
\psi_n(x) \to \psi_n(x)+\e F_n(\psi(x),\pa_\mu \psi(x), \ldots).
\eea
The change in ${\cal L}$ induced thereby is
\bea
\d{\cal L}&=&e \sum_n\left[\frac{\pa \cal L}{\pa \psi_n}F_n +\frac{\pa \cal L}{\pa(\pa_\mu \psi_n)} \pa_\mu F_n + \cdots \right] \nn
& =& \e\sum_n\left[\frac{\pa \cal L}{\pa \psi_n} -\pa_\mu \left(\frac{\pa \cal L}{\pa(\pa_\mu \psi_n)} \right)+ \cdots \right] F_n +\mbox{ total derivative terms}.
\eea
Thus the term $\d{\cal L}=\e \pa{\cal L}/\pa\gamma_0$ in the Lagrangian can be removed by a redefinition of the fields if and only if the term is proportional to the Euler-Lagrange equations. Such an operator is called {\it redundant}, and its coupling {\it inessential}.
As we have seen, operators such as $\bar R$ and $\bar R_{\mu\nu}$ in pure gravity based on the Einstein-Hilbert action are redundant operators, and their couplings are inessential.

We comment on the perturbative renormalizability of the Einstein-Hilbert theory at the two-loop level. 
In this case, the two-loop counter-Lagrangian is given by~\cite{Goroff:1985sz,Goroff:1985th,vandeVen:1991gw}
\begin{align}
\Delta \mathcal L\Big|^\text{2-loop}_\text{EH} 
=-\frac{\sqrt{\bar g}}{(4\pi)^4\varepsilon}\frac{209}{2880} \bar C_{\mu\nu}{}^{\kappa\lambda}\bar C_{\kappa\lambda}{}^{\rho\sigma}\bar C_{\rho\sigma}{}^{\mu\nu}.
\end{align}
This is known as the Goroff-Sagnotti term given by the Weyl tensor $\bar C_{\mu\nu}{}^{\rho\sigma}=\bar R_{\mu\nu}{}^{\rho\sigma}+ \frac{1}{2}(\bar R_{\mu}{}^\sigma \delta_{\nu}^\rho -\bar R_{\mu}{}^\rho \delta_\nu^\sigma +\bar R_{\nu}{}^\rho\delta_\mu^\sigma -\bar R_{\nu}{}^\sigma\delta_\mu^\rho) + \frac{1}{6}\bar R(\delta_{\mu}^\rho\delta_\nu^\sigma -\delta_\mu^\sigma\delta_\nu^\rho)$. 
The two-loop counterterm contains higher derivative operators, e.g., $\bar R_{\mu\nu}{}^{\kappa\lambda}\bar R_{\kappa\lambda}{}^{\rho\sigma}\bar R_{\rho\sigma}{}^{\mu\nu}$ which cannot be removed even if the equation of motion for the background metric field is used. Hence, the perturbative renormalizablity of the Einstein-Hilbert action holds only at the one-loop level.  

Now, we turn to the argument on the scalar-tensor gravity.
In Ref.~\cite{tHooft:1974toh}, 't Hooft and Veltmann have also investigated the perturbative renormalizability of the action
\begin{align}
S_\text{ST} = \int \df^dx \sqrt{g}\left[ -R
- \frac{1}{2}g^{\mu\nu} \p_\mu \phi \p_\nu \phi
\right] ,
\label{eq:simplest SG theory}
\end{align}
where $\phi$ is a real singlet scalar field. We split this as $\phi=\bp+\vp$ where $\vp$ stands for the fluctuation.
The gauge-fixing and ghost terms are given respectively by 
\begin{align}
S_\text{gf} &=
\int_x\sqrt{\bg}\Bigg[\frac{1}{2}\Big(\bnabla_\mu h^{\mu\nu} -\frac{1}{2} \bnabla^\nu h \Big)^2
 + \Big(\bnabla_\mu h^{\mu\nu} -\frac{1}{2} \bnabla^\nu h \Big)(\vp \pa_\nu \bar\phi )
+ \frac{1}{2} (\vp \pa_\mu \bp )^2 \Bigg],
\\
S_\text{gh} &=  \int_x\sqrt{\bg} \,b_\mu \left[ -\d^\mu_\nu \bar{\square} + \pa^\mu\bp \pa_\nu \bar\phi + \br^\mu{}_\nu \right]c^\nu\,.
\end{align}
After eliminating the Riemann tensor squared by using \Cref{eq:Gauss-Bonnet term}, the counter-Lagrangian at the one-loop level was found to be
\begin{align}
\Delta \mathcal L\Big|^\text{1-loop}_\text{ST} =\frac{\sqrt{\bar g}}{8\pi^2\varepsilon}\left( \frac{9}{720}\bar R^2 + \frac{43}{120}\bar R_{\mu\nu}\bar R^{\mu\nu}
+ \frac{1}{2}(\bar g^{\mu\nu}\p_\mu \bar\phi  \p_\nu \bar\phi)^2
-\frac{1}{12}\bar R(\bar g^{\mu\nu}\p_\mu \bar\phi \p_\nu \bar\phi)
+ 2(\bar \square \bar\phi)^2
\right).
\label{eq:counter Lagrangian SG}
\end{align}
The first-order functional derivatives with respect to $\bar g_{\mu\nu}$ and $\bar\phi$ give the equations of motion for the metric and the scalar field as
\begin{align}
&0=\frac{1}{\sqrt{\bar g}}\frac{\delta S_\text{ST}}{\delta \bar\phi}
= \bar\square \bar\phi \,,
\label{eq:EoMforscalarfield}
\\[1ex]
&0=\frac{1}{\sqrt{\bar g}}\frac{\delta S_\text{ST}}{\delta \bar g^{\mu\nu}}
= -\left(\bar R_{\mu\nu}
-\frac12 \bar g_{\mu\nu}\bar R\right)
-\frac12 \p_\mu  \bp \p_\nu \bp
+\frac14 \bar g_{\mu\nu} \left(\bar g^{\rho\lambda}
\p_\rho \bp \p_\lambda \bp \right).
\label{eq:EoMformetric}
\end{align}
Taking the trace for \Cref{eq:EoMformetric}, we obtain
\begin{align}
\bar R =-\frac{1}{2}\bar g^{\mu\nu} \p_\mu \bar \phi \p_\nu \bar\phi.
\label{eq:EoM for R}
\end{align}
Substituting this into \Cref{eq:EoMformetric} yields
\begin{align}
\bar R_{\mu\nu} =-\frac{1}{2} \p_\mu  \bar\phi \p_\nu \bar\phi.
\label{eq:EoM for Rmunu}
\end{align}
Using the relations~\Cref{eq:EoMforscalarfield,eq:EoM for R,eq:EoM for Rmunu}, the counter-Lagrangian~\labelcref{eq:counter Lagrangian SG} is reduced to
\begin{align}
\Delta \mathcal L\Big|^\text{1-loop}_\text{ST} =\frac{\sqrt{\bar g}}{8\pi^2\varepsilon} \frac{203}{80}\bar R^2 .
\end{align}
In contrast to the pure-gravity case, the higher-dimensional operator cannot be eliminated by the equations of motion, so that the scalar-tensor theory~\labelcref{eq:simplest SG theory} is not perturbatively renormalizable even at the one-loop level. However, the situation may be different in the presence of the cosmological constant, which is also considered in Ref.~\cite{Solodukhin:2015ypa}.

\subsection{Higher derivative gravity}
\label{higher}

It is known that if we include curvature square terms in the action
\begin{align}
S_\text{HD} = \int \df^4x \sqrt{g} &\Bigg[
\varrho + a_1 R + a_2R^2 + b_2 C_{\mu\nu}{}^{\rho\sigma}C_{\rho\sigma}{}^{\mu\nu}
+ c \mathfrak{G}	
\Bigg],
\end{align}
where $a_1=-1/(16\pi G_N)$, the theory is perturbatively renormalizable~\cite{Stelle:1976gc}.
However, this theory contains a ghost and is not unitary. 
More specifically, the propagators of graviton (traceless-transverse spin-2 mode) and the conformal mode (spin-0 mode) at the tree level are given by
\begin{align}
[G_g(p)]^{\mu\nu}{}_{\rho\sigma} &= \left[\frac{1}{p^2} - \frac{1}{p^2-2b_2/a_1} \right] (P_t)^{\mu\nu}{}_{\rho\sigma}, \\
G_c(p) &= \frac{1}{p^2} - \frac{1}{p^2-3a_2/2a_1},
\end{align}
where we set $\varrho=0$ and $(P_t)^{\mu\nu}{}_{\rho\sigma}$ is the traceless-transverse projection operator.
The second part in each propagator corresponds to a ghost, indicating the violation of unitarity.
Although several proposals~\cite{Anselmi:2017ygm,Anselmi:2018kgz,Donoghue:2019fcb,Mannheim:2020ryw,Donoghue:2021eto,Mannheim:2021oat} attempt to make the ghost contribution harmless at the quantum level in the sense of the Lee-Wick theory~\cite{Lee:1969fy,Lee:1970iw}, some obstacles still remain~\cite{Coleman:1969xz,Nakanishi:1971jj,Nakanishi:1971kn,Nakanishi:1972wx,Nakanishi:1972pt,Lee:1971ix,Kubo:2023lpz,Kubo:2024ysu}.
For this reason, only few people take this theory seriously, at least at the perturbative level.

\subsection{Asymptotic safety scenario}
\label{ass}

Asymptotically safe gravity is based on a nonperturvative formulation that the UV complete theory for quantum gravity in four-dimensional spacetime is constructed at a nontrivial UV fixed point~\cite{Reuter:1996cp,Souma:1999at}.
This scenario has been investigated mainly in the FRG approach.
In general, one assumes the effective action which is spanned by diffeomorphism invariant effective operators:
\begin{align}
\Gamma_k = \int \df^4x \sqrt{g} \left[ \varrho + a_1R + a_2 R^2 +  b_2 C_{\mu\nu}{}^{\rho\sigma}C_{\rho\sigma}{}^{\mu\nu} + a_3 R^3 +  b_3 C_{\mu\nu}{}^{\rho\sigma}C_{\rho\sigma}{}^{\alpha\beta}C_{\alpha\beta}{}^{\mu\nu} + \cdots \right].
\end{align}
The FRG equation gives the coupled flow equations $\p_t \tilde g_j = \beta_j(\{\tilde g_i\})$ for the dimensionless couplings $\{\tilde g_i\}=\{\tilde\varrho,~\tilde a_1,~\tilde a_2,~\tilde b_2,~\tilde a_3,~\tilde b_3,\cdots\}=\{k^{-4}\varrho,~k^{-2} a_1,~ a_2,~ b_2,~k^{2} a_3,~k^{2} b_3,\cdots\}$ where $k$ is a cutoff scale and $\p_t = k\p_k$.
We explore a fixed point $\{\tilde g_{i*}\}$ at which $\beta_j(\{\tilde g_{i*}\})=0$ for all $i,j$.

In the general framework of the RG flow around a nontrivial fixed point $\tilde g_{i*}\neq 0$, operators are primarily classified into two categories: \textit{relevant} and \textit{irrelevant}.
More specifically, performing the Taylor expansion for the beta functions around the fixed point and taking terms up to linear order, the flow equations become
\begin{align}
\p_t (\tilde g_i -\tilde g_i^*) \simeq \frac{\p \beta_i}{\p \tilde g_k}\bigg|_{\tilde g=\tilde g_*} (\tilde g_j -\tilde g_j^*) \equiv T_{ij}(\tilde g_j -\tilde g_j^*) ,
\label{eq:linearized beta}
\end{align}
where $\beta_i(\{\tilde g_{j*}\})=0$ by definition and $\p_t \tilde g_{i*}=0$.
Its solution reads
\begin{align}
\tilde g_i  = \tilde g_i^* + V_{ij}C_j\left(\frac{k}{\Lambda}\right)^{-\theta_j}.
\end{align}
Here $C_j$ are undetermined constants at a reference scale $\Lambda$ in the differential equations, $V$ are the matrix diagonalizing the stability matrix $T$ and $\theta_j= -\, \text{eig}(T)$ are the critical exponents (or scaling dimensions).
Relevant operators are those whose couplings have $\theta_i>0$ and thus UV attractive fixed points but grow toward the infrared (IR) limit and thus are free parameters which typically need to be fixed by experimental data. 
Irrelevant operators are those whose couplings have $\theta_i<0$ and no UV attractive fixed points and so have to be tuned to flow to fixed values, typically 0, in the UV.
To ensure predictivity in a quantum field theory, only a finite number of directions in coupling space should be relevant; the remaining (irrelevant) directions must be sent to the UV fixed values.
Therefore, it is very important to identify the necessary relevant operators.

In practice, however, deriving the flow equations on general backgrounds is a highly arduous task due to complicated tensorial structures; see e.g.,~\cite{Barth:1983hb,deBerredoPeixoto:2004if,Ohta:2013uca,Ohta:2015zwa}.
For this reason, in many earlier studies, it has been supposed that the effective action is spanned only by the polynomial of $R$, i.e., $\Gamma_k= \int\df^4x\sqrt{g}f(R)$ where $f(R)=\varrho + a_1 R +a_2 R^2 +\cdots$ and the spacetime background is a maximally symmetric (de Sitter) space in which both Ricci tensor squared and Riemann tensor squared are proportional to Ricci scalar squared,~see e.g., \cite{Codello:2007bd,Machado:2007ea,Benedetti:2012dx,Falls:2013bv,Benedetti:2013jk,Falls:2014tra,Falls:2017lst,Falls:2018ylp,DeBrito:2018hur}.
It has been explicitly shown in \cite{Sen:2021ffc,Kawai:2023rgy} that there are four relevant operators at the Gaussian (perturbative) fixed point for the quadratic curvature terms in the gravity with the truncated action $\Gamma_k=\int \df^4x \sqrt{g} [\varrho + a_1 R + a_2R^2 + b C_{\mu\nu}{}^{\rho\sigma}C_{\rho\sigma}{}^{\mu\nu}+ c \mathfrak{G}]$, while some other works~\cite{Benedetti:2009rx,Benedetti:2009gn,Hamada:2017rvn} have studied the asymptotic safety scenario within such a truncated effective action in an Einstein space background.
These studies suggest that there are only three relevant directions at the nontrivial fixed point.\footnote{
We note that the results discussed here were obtained using the so-called background field approximation~\cite{Reuter:1997gx} within the FRG framework.
Another computational scheme is the fluctuation approach~\cite{Pawlowski:2020qer,Pawlowski:2023gym}. Notably, this approach also indicates the existence of three relevant directions in pure gravity~\cite{Christiansen:2016sjn,Denz:2016qks}. For further investigations of gravity-matter systems within the fluctuation approach, see Refs.~\cite{Meibohm:2015twa,Christiansen:2017cxa,Christiansen:2017bsy,Pastor-Gutierrez:2022nki}.
See also \cite{Bosma:2019aiu,Knorr:2019atm,Knorr:2022dsx} for the form-factor approach in which the truncated effective action is given by $\Gamma_k = \int \df^4x \sqrt{g} \left[ \varrho + a_1R + R W_{R^2}(\square)R + C_{\mu\nu}{}^{\rho\sigma}W_{C^2}(\square)C_{\rho\sigma}{}^{\mu\nu} \right]$ with form factors $W_{R^2}(\square)$ and $W_{C^2}(\square)$.
}

However, in specific spacetime backgrounds such as de Sitter or Einstein spaces, the Ricci tensor is proportional to the Ricci scalar, making it impossible to distinguish between $\bar R_{\mu\nu}\bar R^{\mu\nu}$ and $\bar R^2$.
Indeed it has been further confirmed on more general backgrounds that there are nontrivial fixed points in the couplings of higher derivative terms~\cite{Falls:2020qhj,Sen:2021ffc} with three relevant directions.
The result of Ref.~\cite{Falls:2020qhj} is obtained only in the limited case of a finite order in $1/G_N$ expansion.
The full order calculation in $1/G_N$ on a general background was given in Refs.~\cite{Sen:2021ffc,Kawai:2023rgy}. It is suggested in~\cite{Sen:2021ffc} there are three relevant directions at a nontrivial fixed point, while doubt is cast on the existence of nontrivial fixed points in~\cite{Kawai:2023rgy}.
Even if they exist, it is still a big challenge to show that the ghosts do not make any harm in the quadratic curvature theory.

In light of this, we now turn to the essential RG approach, which excludes higher-derivative and higher-curvature terms. As a result, we need not be concerned with the presence of ghosts. The theory consisting solely of the Einstein-Hilbert term and a cosmological constant was previously studied in~\cite{Baldazzi:2021orb}. Here, we extend this analysis by coupling gravity to a scalar field. We note that gravity coupled to a shift-symmetric scalar field has been investigated within the essential RG framework in~\cite{Knorr:2022ilz}. In contrast, our analysis does not assume shift symmetry for the scalar field.

\section{Functional RG for scalar-tensor theory}
\label{FRGST}

In this section, we investigate the scalar-tensor theory using the FRG approach.
We begin by presenting the effective action, from which we compute the Hessians---i.e., the two-point functions obtained via second functional derivatives.
We then derive the beta functions using the standard FRG equation.

\subsection{Scalar-tensor theory}
\label{scalar-tensor}

Let us consider a scalar-tensor theory whose action in Euclidean spacetimes is given by
\begin{align}
S_g = \int \df^4x \sqrt{g}\left[ \varrho - \frac{1}{\kappa}R
- \frac{1}{2}g^{\mu\nu} \p_\mu \phi \p_\nu \phi +\frac12 m^2 \phi^2 \right] ,
\label{eq:action for ST theory}
\end{align}
where
\bea
\kappa = 16\pi G_N
\eea
is the gravitational coupling.

We use the FRG method to study renormalizability of the scalar-tensor theory~\labelcref{eq:action for ST theory}. 
In this work, we employ the Wetterich equation~\cite{Wetterich:1992yh} which is formulated
as a functional differential equation
for the one-particle irreducible action (or simply effective effective)
$\Gamma_k$. Its explicit forms will be shown later.
See Refs.~\cite{Reuter:1993kw,Morris:1998da,Berges:2000ew,Aoki:2000wm,Bagnuls:2000ae,Polonyi:2001se,Pawlowski:2005xe,Gies:2006wv,Delamotte:2007pf,Sonoda:2007av,Igarashi:2009tj,Rosten:2010vm} for review.

Our ansatz for the effective action is
\begin{align}
\Gamma_k^\text{SG}=\int_x \sqrt{g}\left[
\varrho
 - \frac{1}{\kappa} R
- \frac{1}{2} g^{\mu\nu} \pa_\mu \phi\,  \pa_\nu  \phi
+ \frac{1}{2} m^2 \phi^2
\right] ,
\label{eq:starting effect action}
\end{align}
in which $\int_x=\int \df^4x$, $\varrho$ is the cosmological constant, and we have suppressed the explicit dependence of the cutoff scale in the coupling $G_N$.

We calculate fluctuations of the metric and scalar fields around fixed background fields. More specifically,
we split these fields ($g_{\mu\nu}$, $\phi$) into background fields ($\bar g_{\mu\nu}$, $\bar\phi$) and fluctuations
($h_{\mu\nu}$, $\varphi$) as
\begin{align}
&g_{\mu\nu} = \bar g_{\mu\nu} + h_{\mu\nu}\,,
&\phi = \bar\phi + \varphi\,.
\end{align}
To derive the RG equations for couplings in the effective action~\labelcref{eq:starting effect action},
we need to evaluate the second-order variations in terms of the fluctuation fields and the Hessians, i.e.,
the second-order functional derivatives with respect to the fluctuation fields. Its explicit forms are exhibited
in \Cref{appsec:Variations}.

\subsection{Hessians}
\label{hessian}

The second-order terms in the fluctuation fields of the effective action for the scalar-gravity sector $\Gamma_k^\text{SG}$
take the form as
\begin{align}
\Gamma_k^{\text{SG}}
&= \int_x\sqrt{\bar g} \Bigg[\frac{1}{2} h^{\mu\nu}
{\mathcal M}_{\mu\nu\alpha\beta} h^{\a\b} 
+\frac{1}{2} \vp \left( \bar\square + m^2 \right) \vp 
+\left( h^{\mu\nu}-\frac{1}{2} h \bg^{\mu\nu}  \right) \pa_\mu \bp \pa_\nu \vp
\Bigg]\,,
\label{hess1}
\end{align}
where $h=\bar g^{\mu\nu}h_{\mu\nu}$, and the operator $\mathcal M$ is given in \Cref{appeq:M} in \Cref{appsubsubsec:Hessian without gauge-fixing}.
Note that $\bar\square \varphi =\bar\nabla_\mu \p^\mu \varphi$.

Now, we introduce the gauge-fixing for the gravitational fluctuation field.
To get the gauge-fixing and FP ghost terms simultaneously, we consider the BRST transformation~\cite{Kugo:1981hm,Ohta:2020bgz}:
\begin{align}
& \d_B h_{\mu\nu}=-\d\la(\bar g_{\rho\nu}\bar\nabla_\mu c^\rho + g_{\rho\mu}\bar\nabla_\nu c^\rho),
&
&\d_B \varphi =-\d\la\, c^\rho \pa_\rho \varphi, \nn
& \d_B c^\mu = -\d\la\, c^\rho \bar\nabla_\rho c^\mu, \qquad
\d_B b_\mu =i\d\la B_\mu, &
&\d_B B_\mu =0,
\end{align}
where $\d\la$ is an anticommuting parameter, and $c^\mu$, $b_\mu$, and $B_\mu$ are the FP ghost,
antighost and an auxiliary field, respectively.
The gauge-fixing and ghost action is then given by
\bea
\Gamma_\text{GF+FP} 
&=& -i \int_x\sqrt{\bg} \d_B \left[ b_\mu\left(\chi^\mu-\frac{\kappa}{2} B^\mu \right) \right]/\d\la \nn
&=&\int_x\sqrt{\bg} \Bigg[ B_\mu \chi^\mu -\frac{\kappa}{2} B_\mu B^\mu \Bigg] + \Gamma_\text{FP}\nn
&\simeq&
\int_x\sqrt{\bg} \Bigg[\frac{1}{2\kappa} \chi_\mu \chi^\mu  \Bigg] + \Gamma_\text{FP},
\label{gf}
\eea
where in the last equality the auxiliary field $B_\mu$ is integrated out within the path integral formalism.
The gauge parameter would run under the RG flow, but its value does not affect any physical quantities. So here we ignore its running for simplicity.
We would like to choose an appropriate gauge-fixing condition such that nonminimal terms, e.g.,
$\bar\nabla_\mu \bar\nabla_\nu$, in ${\cal M}$ and ${\cal F}$ are eliminated. To this end, we take
\bea
\chi_\mu &=& \nabla^\nu h_{\mu\nu} -\frac{1}{2} \bnabla_\mu h + \kappa(\vp \pa_\mu \bp).
\label{gf2}
\eea
Since the ghost fields are all quantum, we can replace the other fields involved in the FP terms by their expectation
values at the one-loop level. This gives the ghost term as
\begin{align}
\Gamma_\text{FP}
&= -i \int_x\sqrt{\bg}\, b_\mu \left[ \d^\mu_\nu (-\bar{\square} )+\kappa \pa^\mu\bp \pa_\nu \bar\phi
+\br^\mu{}_\nu \right]c^\nu\nn
&\equiv i \int_x\sqrt{\bg}\, b_\mu \Pi^{\mu}{}_\nu  c^\nu.
\end{align}
The gauge-fixing term is given by
\begin{align}
\Gamma_\text{GF}
&=\int_x\sqrt{\bg} \left[ \frac{1}{2\kappa} \chi_\mu \chi^\mu \right]\nn
&=\int_x\sqrt{\bg}\Bigg[\frac{1}{2\kappa}\Big(\bnabla_\mu h^{\mu\nu} -\frac{1}{2} \bnabla^\nu h \Big)^2
 + \Big(\bnabla_\mu h^{\mu\nu} -\frac{1}{2} \bnabla^\nu h \Big) \vp \pa_\nu \bar\phi
+ \frac{1}{2Z_N} (\vp \pa_\mu \bp)^2 \Bigg] .
 \label{eq:gauge fixing}
\end{align}

Together with \Cref{eq:gauge fixing}, we write the effective action up to the second-order variation~\labelcref{hess1} on the field basis in the form
\bea
\Gamma_k
\simeq \bar\Gamma_k[\bar g_{\mu\nu},\bar\phi]
+ \frac{1}{2}\int_x\sqrt{\bar g}\, (h^{\mu\nu}~ \varphi)
\left({\mathbb K}(-\bar\square) +{\mathbb U} \right)
\pmat{h^{\a\b}\\ \varphi}.
\eea
Note that the linear order term of the fluctuation fields vanishes by imposing the equations of motion.
The first term $\bar\Gamma_k$ is \Cref{eq:starting effect action} with the replacement of the fields
($g_{\mu\nu}$, $\phi$) to the background ones ($\bar g_{\mu\nu}$, $\bar\phi$). 
Here, we have defined 
\bea
{\mathbb K} = \pmat{
{\mathbb K}_{\mu\nu\alpha\beta} & 0 \\
0 & - {\bf 1} }, \qquad
{\mathbb U} = \pmat{
{\mathbb U}_{\mu\nu\alpha\beta} & {\mathbb U}_{\mu\nu}\\
{\mathbb U}_{\alpha\beta} & {\mathbb U}_{\cdot \cdot}} ,
\eea
with
\begin{align}
&{\mathbb K}_{\mu\nu\alpha\beta} = \frac{1}{\kappa}\bigg(\frac{1}{2} \bar g_{\mu\alpha} \bar g_{\nu\beta}
-\frac14 \bg_{\mu\nu}\bg_{\a\b} \bigg), \nn
&{\mathbb U}_{\mu\nu\alpha\beta} = \frac{1}{\kappa} \Big[\bg_{\mu\nu}\br_{\a\b}-\br_{\mu\a}\bg_{\nu\b}-\br_{\mu\a\nu\b}\Big]
- \frac{1}{2}\left(-\frac{2}{\kappa} \br+2\varrho + m^2\bar\phi^2 -\bg^{\rho\s}\pa_\rho\bp \pa_\s\bp \right) \times
\nn
&\phantom{{\mathbb U}_{\mu\nu\alpha\beta} =}
\left(\frac{1}{2}\bg_{\mu\a}\bg_{\nu\b}
 - \frac{1}{4}\bg_{\mu\nu}\bg_{\a\b} \right)
-\bg_{\nu\b}\pa_\mu\bp \pa_\a\bp+\frac{1}{2}\bg_{\mu\nu}\pa_\a\bp \pa_\b \bp,
\nn
&{\mathbb U}_{\mu\nu} = - \bnabla_\nu \bnabla_\mu \bp +\frac12 \bar\square \bp \bg_{\mu\nu}+ \frac{1}{4}m^2 \bar\phi \bg_{\mu\nu}, 
\nn
&{\mathbb U}_{\a\b} = -\bnabla_\a \bnabla_\b\bp+\frac{1}{2}\bar\square \bp \bg_{\a\b} + \frac{1}{4}m^2 \bar\phi \bg_{\a\b}, 
\nn
&{\mathbb U}_{\cdot \cdot} = \kappa(\pa_\rho\bp)^2 + m^2,
\label{finalhessian}
\end{align}
where symmetrization in $(\mu,\nu)$ and $(\a,\b)$ should be understood.
Note that the identity matrix in the field basis $(h_{\mu\nu}, \varphi)$ is given by
\begin{align}
&{\bf 1}=\pmat{
\d^{\a\b}_{\mu\nu}  & 0\\
0 & 1
},&
&\d^{\a\b}_{\mu\nu} \equiv \frac{1}{2}(\d^\a_\mu \d^\b_\nu+\d^\a_{\nu} \d^\b_\mu),
\label{eq:the identity matrix}
\end{align}
and its trace yields $\tr[{\bf 1}]= \d^{\a\b}_{\mu\nu} (\delta^\mu_\alpha \delta^\nu_\beta) +1=11$ corresponding
to the degrees of freedom of a symmetric tensor and a scalar field.
We then rewrite the Hessian as
\bea
\mathcal H &=& {\mathbb K} (-\bar \square +{\mathbb W}),
\label{eq:H kernel}
\eea
with ${\mathbb W} ={\mathbb K}^{-1} {\mathbb U}$.
Here the inverse of the coefficient matrix $\mathbb K$ of $-\bar\square$ is given by
\bea
\mathbb K^{-1}=
\pmat{ \frac{1}{Z_N} \left(\bg_{\mu\a}\bg_{\nu\b}+\bg_{\nu\a}\bg_{\mu\b}-\bg_{\mu\nu}\bg_{\a\b} \right) & 0 \\
0 & -1 }.
\eea
Then we define
\bea
\mathcal D\equiv {\mathbb K}^{-1}\mathcal H =-\bar \square  +{\mathbb W}.
\eea

\subsection{Standard RG equation}
\label{standard}

Let us first employ the standard form of the RG equation~\cite{Wetterich:1992yh} (see also \cite{Ellwanger:1993mw,Morris:1993qb}),
\begin{align}
\p_t \Gamma_k = \frac{1}{2}\Tr \left[ \left( \Gamma_k^{(2)} + \mathcal R_k \right)^{-1} \p_t \mathcal R_k \right]
= \frac{1}{2}\left\{\left[ \left( \Gamma_k^{(2)} + \mathcal R_k \right)^{-1} \right] ^{AB} \left( \delta^C_B \p_t \right) [\mathcal R_k]_{CA} \right\}.
\end{align}
For the current setup, we have
\begin{align}
\pa_t \G_k  = \frac{1}{2} \Tr\left[\frac{\pa_t \mathcal R_k(\mathcal H)}{\G_k^{(2)}(\mathcal H)+\mathcal R_k(\mathcal H)}\right] 
- \Tr\left[\frac{\pa_t \mathcal R_k(\Pi)}{\G_\text{FP}^{(2)}(\Pi)+\mathcal R_k(\Pi)}\right].
\label{eq:flow equation for SG}
\end{align}
We compute the RG kernels by using the heat kernel method. 
Its formula and the detailed processes for the computations are summarized in \Cref{appsec:Heat kernel expansion}.
Here the first term on the right-hand side of \Cref{eq:flow equation for SG} is computed by the heat kernel expansion as
\begin{align}
&\frac{1}{2} \Tr\left[\frac{\pa_t (\mathcal {\mathbb K} R_k(\mathcal D))}{{\mathbb K}\mathcal D +\mathcal {\mathbb K} R_k(\mathcal D)}\right]
\simeq \frac{1}{2} \Tr\left[\frac{\pa_t R_k(\mathcal D)}{\mathcal D  +R_k(\mathcal D)}\right]\nn
&=\frac{1}{2(4\pi)^2}
\int_x \sqrt{\bar g}
\left[ Q_2 \tr[{\bf b}_0(\mathcal D)]  + Q_1\tr[{\bf b}_2(\mathcal D)] + Q_0\tr[{\bf b}_4(\mathcal D)]+\cdots   \right],
\end{align}
where $\mathcal R_k(\mathcal H)=\mathbb K R_k(\mathcal D)$ and we have neglected the term with the field anomalous dimensions $\sim \p_t \mathbb K/\mathbb K$ on the numerator and have defined the following threshold functions by performing the inverse Laplace transformation for $s$: 
\begin{align}
Q_2&=\int_0^\infty \df z\,z \frac{\p_t R_k(z)}{z+R_k(z)}
=k^4,\\
Q_1&=\int_0^\infty \df z \frac{\p_t R_k(z)}{z+R_k(z)}
=2k^2,\\
Q_0&= \frac{\p_t R_k(z)}{z+R_k(z)}\bigg|_{z=0}
=2,
\end{align}
where in the second equality, we have employed the Litim-type cutoff function~\cite{Litim:2001up}
\begin{align}
R_k(z) = (k^2-z)\theta(z-k^2).
\end{align}
The heat kernel coefficients $\tr[{\bf b}_i(\mathcal D)]$ are given in \Cref{eqApp:heat1,eqApp:heat2,eq:D4 hkc} in \Cref{appsec:Heat kernel expansion}.

In the same manner, we compute the ghost contribution
\begin{align}
&- \Tr\left[\frac{\pa_t \mathcal R_k(\Pi)}{\G_\text{FP}^{(2)}(\Pi)+\mathcal R_k(\Pi)}\right]\nn
&=-\frac{1}{(4\pi)^2}\int_x \sqrt{\bar g}
\left[ Q_2 \tr[{\bf b}_0(\Pi)]  + Q_1\tr[{\bf b}_2(\Pi)] + Q_0\tr[{\bf b}_4(\Pi)]+\cdots   \right],
\end{align}
with the heat kernel coefficients \labelcref{eqapp:ghost heat1,eqapp:ghost heat2,eqapp:ghost heat3}.

Collecting all these contributions, we find the RG equation for the effective action
\begin{align}
&\p_t\Gamma_k = \frac{1}{2(4\pi)^2}\int_x\sqrt{\bar g}\Bigg[
3 Q_2 + ( m^2+10 \kappa\varrho) Q_1 +\left(\frac{m^4}{2} +5\kappa^2 \varrho^2 \right) Q_0 \nn
&\qquad + \left\{-\frac{15}{2} Q_1 + \left( \frac{m^2}{6} -\frac{13}{3}\kappa \varrho \right) Q_0\right\} \br + \frac{49}{240} Q_0\br^2  + \frac{283}{60} Q_0 S_{\mu\nu}S^{\mu\nu}
 -\kappa Q_1  \p_\mu \bar\phi \p^\mu \bar\phi \nn
&\qquad + \left\{ 5\kappa Q_1 + \left(\frac{m^2}{2}\kappa+5\kappa^2\varrho\right) Q_0 \right\} m^2\bar\phi^2 +\frac{5}{4}m^4\kappa^2 Q_0 \bp^4 -\frac{13}{6}m^2\kappa Q_0 \bar\phi^2 \br \nn
&\qquad
-\frac{1}{6} \kappa Q_0 \bar R (\p_\mu\bar\phi \p^\mu\bar\phi) 
+ 4 \kappa Q_0 S^{\mu\nu}\p_\mu\bar\phi\p_\nu\bar\phi
- \kappa Q_0 (\bar\square \bar\phi)^2
+\frac{7}{4}\kappa^2 Q_0 (\p_\mu\bar\phi\p^\mu\bar\phi)^2
\Bigg],
\label{divergentt}
\end{align}
where $\bar S_{\mu\nu}=\bar R_{\mu\nu}-\frac14 \bar g_{\mu\nu}\bar R$ is the traceless Ricci tensor and we have ignored total derivative terms such as $\bar\square \bar R$ and $\bar\square (\p_\mu\bar\phi \p^\mu\bar\phi)$.
We see that there are several divergent terms like $\br^2$ and $\br(\pa_\mu\bp \pa^\mu\bp)$
which is not present on the left-hand side, and so the theory is not renormalizable. However we may try to remove these by the field redefinitions, making the theory renormalizable.

\section{Essential RG equation}
\label{essential}

We have seen in \Cref{'tHooft-Veltman} that we can remove various divergences by the field redefinitions if the terms are proportional to the field equations.
The essential RG seeks for the field redefinitions that remove new divergent terms that cannot be dealt with by renormalization. In this way, this method may give a nonperturbatively renormalizable theory.

The essential RG equation is given by~\cite{Pawlowski:2005xe,Baldazzi:2021ydj,Baldazzi:2021orb} (see also~\cite{Wetterich:2024uub})
\begin{align}
\left( \pa_t +\Psi^A \frac{\d}{\d\Phi^A}\right) \Gamma_k= \frac{1}{2}\left\{ \left[\left(\G_k^{(2)}+\mathcal R_k \right)^{-1}\right]^{AB} \left( \d^C_B \pa_t +2\frac{\d\Psi^C}{\d\Phi^B}\right) [\mathcal R_k]_{CA} \right\},
\label{rgf}
\end{align}
where $\Psi^A$ are called RG kernels as defined below. We choose 
the redundant operator basis as
\bea
\left\{ \bg_{\mu\nu},\quad \br \bg_{\mu\nu},\quad \bar S_{\mu\nu},\quad \br^2 \bg_{\mu\nu},\quad \br \bar S_{\mu\nu},\quad \bp^2 \bg_{\mu\nu}, \quad \p_\mu \bp \p_\nu \bp  \right\},~~~
\{\bp, \quad \bar{\square} \bp, \quad \bp^3 \}.
\eea
With this basis, the RG kernel is defined by
\bea
&&\Psi_{\mu\nu}^g = \gamma_g \bg_{\mu\nu}+\gamma_R \br \bg_{\mu\nu}
+\gamma_{S} \bar S_{\mu\nu}+\gamma_{R^2} \br^2 \bg_{\mu\nu}
+ \gamma_{S^2} \bar S_\mu{}^\s \bar S_{\s\nu}+\gamma_{g\phi} \bp^2 \bg_{\mu\nu}+ \gamma_{\p\phi\p\phi} \p_\mu \bp \p_\nu\bp\,, \nn[1ex]
&& \Psi^\phi =\gamma_\phi \bp +\gamma_{\square \phi} \bar\square \bp +\gamma_{\phi^3} \bp^3 \,.
\eea
Combining the RG kernel and the field equations collected in appendix~\ref{eom}, we find
\begin{align}
& \Psi^A \frac{\d \G_k}{\d \Phi^A}
 =
\Psi^g_{\mu\nu}\frac{\delta\Gamma_k}{\delta \bar g_{\mu\nu}} + \Psi^\phi
\frac{\delta\Gamma_k}{\delta \bar \phi}\nn
&=\int_x \sqrt{\bg} \Big[ 2 \gamma_g \varrho +\Big( -\frac{\gamma_g}{\kappa} +
2\varrho \gamma_R \Big) \br +\Big( -\frac{\gamma_R}{\kappa} + 2\gamma_{R^2} \varrho
\Big) \br^2 +\Big(-\frac{\gamma_S}{\kappa} +
\frac{\gamma_{S^2}}{2} \varrho \Big) \bar S^{\mu\nu} \bar S_{\mu\nu}
\nn
&+\left( \gamma_g m^2+\gamma_\phi m^2 +2\gamma_{g\phi} \varrho \right) \bp^2 
+ \left(-\frac{\gamma_g}{2} + \frac{\gamma_{\p\phi\p\phi}}{2}\varrho \right)(\pa_\mu \bp)^2 
+ (\gamma_\phi +\gamma_{\square\phi}m^2)\bp\bar\square\bp \nn
& +\gamma_{\square\phi}(\bar \square \bp)^2 + \left( -\frac{\gamma_S}{2} - \frac{\gamma_{\p\phi\p\phi}}{\kappa} \right)\bar S^{\mu\nu}\pa_\mu \bp \pa_\nu \bp
+ \left( \gamma_R m^2 -\frac{\gamma_{g\phi}}{\kappa}\right) \bp^2 \br + (\gamma_{g\phi}+\gamma_{\phi^3}) m^2 \bp^4\nn
& + \left(-\frac{\gamma_R}{2} - \frac{1}{4}\frac{\gamma_{\p\phi\p\phi}}{\kappa} \right)(\pa_\mu \bp)^2 \br
- \frac{1}{2} \gamma_{\p\phi\p\phi} (\p_\mu \bp \p^\mu\bp)^2
+\left(-\frac{\gamma_{g\phi}}{2}+\frac14 \gamma_{\pa\phi\pa\phi}m^2 \right)\bp^2 (\pa_\mu\bp)^2 \nn
& -\frac12 \gamma_{R^2} \br^2(\pa_\mu\bp)^2+\gamma_{R^2} m^2 \br^2 \bp^2
+\frac{\gamma_{S^2}}{4}m^2\bp^2 \bar S_{\mu\nu}^2
+\gamma_{\phi^3} m^2 \bp^3 \bar\square \bp,
\label{divergentt2}
\end{align}
where we have dropped some terms of higher derivatives more than 4 in the sprit of derivative expansion.
We also find
\begin{align}
&& \hs{-10} \frac{\d\Psi^g_{\mu\nu}}{\d \bar g_{\a\b}}
= \gamma_g \d^{(\a}_{(\mu}\d^{\b)}_{\nu)} +\Big(\gamma_R-\frac14 \gamma_S\Big) \left[
\br \d^{(\a}_{(\mu}\d^{\b)}_{\nu)}-\br^{\a\b}\bar g_{\mu\nu}+ \bg_{\mu\nu} (\bar\nabla^{(\a}\bar\nabla^{\b)}-\bar g^{\a\b}\bar\square) \right] \nn
&& +\gamma_S \Big(-\frac12 \br^\a{}_\mu{}^\b{}_\nu+\br^{(\a}_{(\mu} \d_{\nu)}^{\b)}
-\frac12 \bar g^{\a\b}\bar\nabla_\mu \bar\nabla_\nu + \d^{(\a}_{(\mu} \bar\nabla_{\nu)}\bar\nabla^{\b)} \Big) \nn
&& + \gamma_{R^2} \Big[2\br \bg_{\mu\nu}(\bar\nabla^\a \bar\nabla^\b -\bar\square \bg^{\a\b}-\br^{\a\b}) +\br^2 \d^{(\a}_{(\mu} \d^{\b)}_{\nu)} \Big] \nn
&& +\gamma_{S^2} \bar S_{\rho(\mu} \Big[ -\bg^{\a\b}\bar\nabla^\rho\bar\nabla_{\nu)}+
\d^\b_{\nu)}\bar\nabla^\rho\bar\nabla^\a +\bg^{\a\rho} \bar\nabla_{\nu)}\bar\nabla^\a-\br^{\a\rho\b}{}_{\nu)}+\br^{\a\rho}\d^\b_{\nu)}+\br^\a{}_{\nu)}\bg^{\b\rho} \nn
&& -\frac12 \br \bg^{\rho(\a}\d_{\nu)}^{\b)}-\frac12\d^\rho_{\nu)}(\bar\nabla^\a \bar\nabla^\b-\bg^{\a\b}\bar\square-\br^{\a\b}) \Big]
+\gamma_{g\phi} \bar\phi \d_{(\mu}^{(\a}\d_{\nu)}^{\b)},
\label{ad1}
\end{align}
and
\begin{align}
\frac{\d\Psi_\phi}{\d \bar\phi} =\gamma_\phi + \gamma_{\square\phi}\bar\square + 3\gamma_{\phi^3}\bar \phi^2,
\label{ad2}
\end{align}
where the identity in the space of symmetric tensor $\d^{\a\b}_{\mu\nu}$ has been defined in \Cref{eq:the identity matrix}.

Ignoring the latter contributions~\labelcref{ad1,ad2}, which are analogous to the anomalous dimensions,\footnote{The reason is that these terms correspond to resummation of multiloop corrections which may be dropped in the one-loop approximation.}
we find the divergent terms may be eliminated by the following conditions: 
\begin{align}
\br \bp^2:& \hs{10} \gamma_R m^2 -\frac{\gamma_{g\phi}}{\kappa} =  -\frac{13}{12(4\pi)^2}m^2\kappa Q_0,  \\[1ex]
\br^2:& \hs{10} -\frac{\gamma_R}{\kappa} +2\gamma_{R^2}\varrho = \frac{49}{480(4\pi)^2}Q_0,\\[1ex]
\bar S_{\mu\nu}^2: & \hs{10} -\frac{\gamma_S}{\kappa} + \frac{\gamma_{S^2}}{2}\varrho = \frac{283}{120(4\pi)^2}Q_0,\\[1ex]
\br (\pa_\mu\bp)^2: & \hs{10} -\frac{\gamma_R}{2} - \frac{1}{4}\frac{\gamma_{\p\phi\p\phi}}{\kappa} = -\frac{1}{12(4\pi)^2}\kappa Q_0,\\[1ex]
\bar S^{\mu\nu} \pa_\mu\bp \pa_\nu \bp: & \hs{10} -\frac{\gamma_S}{2} -\frac{\gamma_{\p\phi\p\phi}}{\kappa} = \frac{2}{(4\pi)^2}\kappa Q_0,\\[1ex]
(\bar\square\bp)^2: & \hs{10} \gamma_{\square\phi} =-\frac{1}{2(4\pi)^2}\kappa Q_0,\\[1ex]
\bp^4:& \hs{10} (\gamma_{g\phi} +\gamma_{\phi^3})m^2 =  \frac{5}{8(4\pi)^2}m^4\kappa^2 Q_0,  \\[1ex]
(\pa_\mu\bp \pa^\mu\bp)^2:& \hs{10} - \frac{1}{2} \gamma_{\p\phi\p\phi} = \frac{7}{8(4\pi)^2}\kappa^2 Q_0 ,
\end{align}
where we have indicated the operators with divergences in the first column, which are canceled by the relations listed.
Solving these relations for $\gamma$'s, we obtain
\begin{align}
&\gamma_{\p\phi\p\phi} = -{\frac{7}{4(4\pi)^2}}\kappa^2 Q_0,~~
\gamma_{\square\phi} = -{\frac{\kappa}{2(4\pi)^2}} Q_0,\quad
\gamma_S = -\frac{\kappa}{2(4\pi)^2} Q_0,\quad
\gamma_R = \frac{25\kappa}{24(4\pi)^2} Q_0, \nn
&\gamma_{S^2} = \frac{223}{60(4\pi)^2\varrho}Q_0, \quad
\gamma_{R^2} = \frac{183}{320(4\pi)^2\varrho}Q_0, \quad
\gamma_{g\phi} = \frac{17m^2\kappa^2}{8(4\pi)^2} Q_0,\quad
\gamma_{\phi^3} = -\frac{3 m^2\kappa^2}{2(4\pi)^2} Q_0.
\end{align}
Thus we find that we can remove the divergent terms in \Cref{divergentt} which are absent in the effective action by the field redefinitions.
Note that this is possible because we have nonzero vacuum energy $\varrho$. Otherwise, we cannot use $\gamma_{S^2}$ to remove a divergence, for example.
This is in sharp contrast to the perturbative results, where the cosmological constant is absent and the $R^2$ term could not be eliminated through the field redefinition (see \Cref{'tHooft-Veltman}). A similar observation was made in Ref.~\cite{Solodukhin:2015ypa}.

However, this process introduces new divergent terms in the RG equation. Those are the last five terms in \Cref{divergentt2}. Actually we can remove these new divergences by further field redefinitions, but these do not affect the coefficients of the Einstein and vacuum energy terms. So, these are not exposed explicitly here. We also note that this is possible again because of the presence of nonzero vacuum energy.
In this way we can systematically move the divergent counterterms to higher derivative terms, which should be dealt with at the next order. These terms are of higher dimensions and are expected to be more irrelevant, like the Goroff-Sagnotti term.

The remaining terms give the RG equations for $\varrho$ and $\kappa$:
\begin{align}
\pa_t \varrho &= -2\gamma_g \varrho +\frac{1}{2(4\pi)^2}
\left[3 k^4 +2(m^2+10 \kappa \varrho)k^2+m^4 +10 \kappa^2 \varrho^2\right],
\label{firstrg}
\\
\pa_t\kappa &= \gamma_g \kappa -2 \gamma_R \varrho \kappa^2 +\frac{1}{2(4\pi)^2}
\left( -15 k^2 + \frac{m^2}{3}-\frac{26}{3}\kappa \varrho \right) \kappa^2.
\label{secondrg}
\end{align}
Here we see that there is an undetermined parameter $\gamma_g$.
This corresponds to the freedom of the wave function renormalization of the metric~\cite{Kawai:2023rgy,Kawai:2024rau}.
In fact, if we make the wave function renormalization $g_{\mu\nu}=Z g_{\mu\nu}'$, the couplings have to scale as
\begin{align}
\varrho'=Z^2 \varrho, \qquad
\kappa' = Z^{-1} \kappa.
\label{eq:rescaled rho and kappa}
\end{align}
Then the cutoff $k$ also has to scale as
\begin{align}
k'^2 =Z k^2.
\label{eq:rescaled scale}
\end{align}
For the infinitesimal transformation $Z=1-\delta Z$, \Cref{eq:rescaled rho and kappa,eq:rescaled scale} imply that
\bea
\d\varrho = -2 \delta Z \,\varrho, \qquad
\d \kappa= \delta Z \,\kappa, \qquad
\d k=-\frac{\delta Z}{2} k.
\eea
By setting $\delta Z=\gamma_g \df t$, we see that the first two equations gives precisely the terms involving $\gamma_g$ in our RG equations~\labelcref{firstrg,secondrg}.

Using this freedom, it is possible to choose $\gamma_g$ such that the loop corrections to the coupling of $\bar R$ cancel:
\begin{align}
\gamma_g = 2 \gamma_R \varrho \kappa - \frac{1}{2(4\pi)^2}
\left( -15 k^2 + \frac{m^2}{3}-\frac{26}{3}\kappa \varrho \right) \kappa,
\end{align}
where $\gamma_R$ is already determined above.
For this choice, the Newton coupling does not run.
Another possibility is to cancel the loop corrections to the cosmological constant:
\begin{align}
2\gamma_g \varrho =- \frac{1}{2(4\pi)^2}
\left[3 k^4 +2(m^2+10 \kappa \varrho)k^2+m^4 +10 \kappa^2 \varrho^2\right),
\end{align}
for which the cosmological constant does not run.

This is due to the dimensional nature of these couplings, and the RG equations~\labelcref{firstrg} and \labelcref{secondrg} do not make much physical sense separately.
On the other hand, we can also make the RG equations invariant under the wave function renormalization of the metric as follows.
We see that $\gamma_g$ drops out if we write down the RG equations for the dimensionless couplings~\cite{Kawai:2023rgy,Kawai:2024rau}:
\begin{align}
\tilde \varrho = \varrho k^{-4}, \qquad
\tilde \kappa = \kappa k^2, \qquad
\tilde m^2 = m^2 k^{-2}.
\end{align}
Substituting all these into \Cref{rgf}, we find
\begin{align}
\p_t \tilde\varrho &= -4\tilde\varrho + \frac{1}{2(4\pi)^2}\left[ 3 + 2 (\tilde m^2+10 \tilde\kappa \tilde\varrho) +\tilde m^4+10\tilde\kappa^2 \tilde\varrho^2 \right],\\
\p_t \tilde \kappa &= 2\tilde \kappa - 2 \gamma_R\, \tilde\varrho\, \tilde\kappa^2 + \frac{1}{2(4\pi)^2}\left( -15 + \frac{\tilde m^2}{3} -\frac{26}{3}\tilde\kappa \tilde \varrho \right) \tilde \kappa^2,\\
\p_t \tilde m^2 &= -2(1+\gamma_\phi)\tilde m^2 -4 \gamma_{g\phi} \tilde\varrho+\frac{1}{(4\pi)^2}\left( 10\tilde\kappa+ \tilde \kappa\tilde m^2+10\tilde \kappa^2\tilde\varrho \right) \tilde m^2.
\end{align}
The factor $\gamma_\phi$, related to the wave function renormalization of $\bp$, should be chosen such that the kinetic term of $\bp$ is correctly normalized:
\bea
\gamma_\phi = \frac{\gamma_{\pa\phi\pa\phi}}{2}\varrho-\gamma_{\square\phi} m^2 +\frac{\kappa}{(4\pi)^2}.
\eea
Using the $\gamma$'s already determined, we find the fixed point with positive $\tilde\kappa$ is
\begin{align}
(\tilde\varrho_*,~\tilde\kappa_*,~\tilde m^2_*)
=(0.005896,~34.265,~0).
\label{eq:fixedpoint value}
\end{align}
Note that the this corresponds to the fixed-point value of the dimensionless Newton coupling $\tilde G_{N*} =\tilde \kappa_*/(16\pi) = 0.682$. 
At this fixed point, the values of $\gamma$ factors are 
\begin{align}
&\gamma_{R*} = 0.452048,&
&\gamma_{g\phi*}=0,&
&\gamma_{R^2*}=1.22837,\nn
&\gamma_{S*}=-0.216983,&
&\gamma_{S^2*}=7.98328,&
&\gamma_{\p\phi\p\phi*}=-26.022,\nn
&\gamma_{\square\phi*}=  -0.216983,&
&\gamma_{\phi^3*}=0,&
&\gamma_{\phi*}=0.21674.
\end{align}
Here some of the $\gamma$ factors---specifically $\gamma_{R^2}$, $\gamma_{S^2}$, and $\gamma_{\partial\phi\,\partial\phi}$, which are associated with higher-dimensional operators---are relatively large compared to others. This may be an artifact of the current truncation. It is therefore expected that including additional higher-dimensional operators in the effective action would reduce these values. In contrast, $\gamma_{R*}$, $\gamma_{g\phi*}$, and $\gamma_{\phi*}$, which contribute to the flow equations of the essential couplings, are small. As a result, they do not significantly alter the critical behavior near the fixed point, as will be shown below.

The stability matrix \labelcref{eq:linearized beta} at the fixed point \labelcref{eq:fixedpoint value} is evaluated as
\begin{align}
T = \pmat{
-1.39178 & 0.000448828 & 0.00633257 \\
-2165.39 & -2.37263 & 1.23914\\
0 & 0 & -0.570517
},
\end{align}
from which the critical exponents are found to be
\begin{align}
&\theta_1 =1.882 + 0.855i,&
&\theta_2 =1.882 - 0.855i,&
&\theta_3 =0.571.
\end{align}
This indicates that all three directions are relevant.
In particular, the critical exponent of the scalar mass parameter becomes smaller than its canonical dimension $2$.
For some scaling solutions, see~\cite{Henz:2013oxa,Henz:2016aoh,Wetterich:2019rsn,Percacci:2014tfa,Labus:2015ska,Bonanno:2025qsc}.
It may appear that the critical exponent for the mass parameter
scaling is much smaller than expected. Several studies, for instance~\cite{Percacci:2003jz, Eichhorn:2020sbo}, have reported that, in gravity-scalar theories, the critical exponent of the scalar mass tends to significantly decrease. Our fixed-point value and the associated critical exponent are consistent with these earlier findings.
This fact implies that the RG scaling of the scalar mass parameter is mild and thus it is a hint for the solution to the gauge hierarchy problem thanks to gravitational fluctuations~\cite{Wetterich:2016uxm}.

\section{Summary}
\label{summary}

In this paper, we have summarized the perturbative on-shell renormalizability of the Einstein gravity~\cite{tHooft:1974toh}, which is equivalent to the renormalizability by the field redefinitions, and the nonrenormalizability of the theory at the two-loop level due to the presence of the Goroff-Sagnotti term~\cite{Goroff:1985sz,Goroff:1985th,vandeVen:1991gw}.
The FRG approach to the Einstein theory with the vacuum energy (equivalent to the cosmological constant) has been shown to be nonperturbatively renormalizable at the two-loop level if the field redefinition is incorporated. 
Dangerous counterterms are redundant operators and may be removed by the field redefinitions, and the two-loop nonrenormalizable term is irrelevant~\cite{Gies:2016con,Baldazzi:2023pep}.

It has been known that the Einstein gravity becomes perturbatively nonrenormalizable already at one loop when coupled to scalar fields on the flat background~\cite{tHooft:1974toh}. We have studied the system with the vacuum energy on the general backgrounds, and find that the counterterms may be removed by the field redefinitions. We have found that the presence of the vacuum energy is quite essential in formulating the nonperturbative renormalizability by the field redefinitions. There may still be possible higher dimensional operators which should be dealt with at the next loop order. But we expect that such terms are either removed by further field redefinitions if they do not involve the Riemann tensor, or if they contain the Riemann tensors, they are irrelevant because the higher the dimensions of the operators are, the more irrelevant the operators are. In fact, we already know that the cubic Riemann-tensor term (the above-mentioned Goroff-Sagnotti term) is irrelevant.

\section*{Acknowledgments}
We thank Kevin Falls for valuable comments.

\appendix
\section{Hessians}
\label{appsec:Variations}

\subsection{Hessian without gauge-fixing}
\label{appsubsubsec:Hessian without gauge-fixing}

The second-order variational term for \Cref{eq:starting effect action} is given by
\begin{align}
\delta^2\Gamma_k^{\text{SG}}
&= \int_x\sqrt{\bar g} \Bigg[\frac{1}{2} h^{\mu\nu} {\mathcal M}_{\mu\nu\alpha\beta} h^{\a\b}
 +\frac{1}{2} \vp \left( \bar\square + m^2 \right)\vp 
+\left( h^{\mu\nu}-\frac{1}{2} h \bg^{\mu\nu}  \right) \pa_\mu \bp \pa_\nu \vp
\Bigg]\,,
\end{align}
with
\begin{align}
&{\mathcal M}_{\mu\nu\alpha\beta}=
- \frac{1}{\kappa}\bigg[\frac{1}{2} \bar g_{\mu\alpha} \bar g_{\nu\beta} \bar\square  - \frac{1}{2} \bar g_{\mu\nu}\bar g_{\alpha\beta}\bar\square - \bar g_{\nu\alpha} \bar \nabla_\mu \bar \nabla_\beta + \bar g_{\mu\nu} \bar \nabla_\alpha \bar \nabla_\beta + \bar R_{\mu\alpha} \bar g_{\nu\beta} - \bar g_{\mu\nu}\bar R_{\alpha \beta}  \nn
&\quad + \bar R_{\mu\alpha\nu\beta} + \frac{1}{4} \bar R ( \bar g_{\mu\nu}\bar g_{\alpha\beta}- 2 g_{\mu\alpha} \bar g_{\nu\beta}) \bigg] +\frac{1}{2} \left( 2\varrho + m^2\bar\phi^2 -\bg^{\la\s}\pa_\la\bar\phi \pa_\s \bar\phi\right) \left( \frac{1}{4}\bar g_{\mu\nu}\bar g_{\alpha\beta} - \frac{1}{2} \bar g_{\mu\alpha} \bar g_{\nu\beta} \right) \nn
&\quad - \bar g_{\nu\b} \p_\mu \bp \p_\a \bp + \frac{1}{2} \bar g_{\mu\nu} \p_\a \bar\phi \p_\b\bp \,,
\label{appeq:M}
\end{align}

\subsection{Hessian with gauge-fixing}
\label{hessiangf}

The second-order variational term for \Cref{eq:starting effect action} with
the gauge-fixing term \labelcref{eq:gauge fixing} is given by
\begin{align}
&\delta^2(\Gamma_k^{\text{SG}}+\Gamma_\text{GF}) 
\nn
&= \int_x\sqrt{\bar g} \Bigg[\frac{1}{2} h^{\mu\nu} \widetilde{\mathcal M}_{\mu\nu\alpha\beta} h^{\a\b} +\frac{1}{2} \vp\bar\square \vp
 +\frac{\kappa}{2} (\pa_\mu\bp)^2\vp^2
-\left( h^{\mu\nu} -\frac{1}{2} h \bg^{\mu\nu} \right) (\bnabla_\mu \bnabla_\nu\bp) \vp
\Bigg] \,,
\label{hess2}
\end{align}
with
\begin{align}
&\widetilde{\mathcal M}_{\mu\nu\alpha\beta}
=
- \frac{1}{\kappa}\bigg[\frac{1}{2} \bar g_{\mu\alpha} \bar g_{\nu\beta} \bar\square
- \frac{1}{4} \bar g_{\mu\nu}\bar g_{\alpha\beta}\bar\square + \bar R_{\mu\alpha} \bar g_{\nu\beta}
- \bar g_{\mu\nu}\bar R_{\alpha \beta} + \bar R_{\mu\alpha\nu\beta} \nn
& \quad + \frac{1}{4} \bar R ( \bar g_{\mu\nu}\bar g_{\alpha\beta}- 2 \bg_{\mu\alpha} \bg_{\nu\beta}) \bigg]
+\frac{1}{2} \left( 2\varrho + m^2\bar\phi^2 -\bg^{\rho\s}\pa_\rho\bar\phi \pa_\s \bar\phi\right)\left(\frac{1}{4}\bar g_{\mu\nu}\bar g_{\alpha\beta} - \frac{1}{2} \bar g_{\mu\alpha} \bar g_{\nu\beta} \right)\nn
& \quad - \bar g_{\nu\b} \p_\mu \bp \p_\a \bp + \frac{1}{2} \bar g_{\mu\nu} \p_\a \bar\phi \p_\b\bp \,.
\end{align}

\section{Heat kernel expansion}
\label{appsec:Heat kernel expansion}

The right-hand side of the RG equation take typically a form of
\begin{align}
W[\Delta] =\frac{\pa_t  R_k(\bar\Delta)}{\bar\Delta + R_k(\bar\Delta)},
\end{align}
with a certain Laplacian $\bar\Delta$.
Now, we deform this RG kernel as follows: 
\begin{align}
\frac{1}{2}  \Tr \,W[\bar\Delta]
&=\frac{1}{2}  \sum_m W[\lambda_m]
=\frac{1}{2}  \sum_m \int_0^\infty \df z\delta(z -\lambda_m)W[z]
\nn
&=\frac{1}{2}\int_0^\infty \df z\,W[z]
\int_{\gamma-i\infty}^{\gamma+i\infty}\frac{\df s}{2\pi i}\,e^{sz}\, \tr [e^{-s \bar\Delta}]\nn
&=
\frac{1}{2}
\int_x \sqrt{\bar g}
\int_0^\infty \df z \,W[z]
\int_{\gamma-i\infty}^{\gamma+i\infty}\frac{\df s}{2\pi i} \,e^{sz}
\left[ \tr[{\bf b}_0(\bar\Delta)]  +\tr[{\bf b}_2(\bar\Delta)]s + \tr[{\bf b}_4(\bar\Delta)]s^4+\cdots   \right]\nn
&=\frac{1}{2}
\int_x \sqrt{\bar g}
\left[ Q_2[W] \tr[{\bf b}_0(\bar\Delta)]  + Q_1[W]\tr[{\bf b}_2(\bar\Delta)]s + Q_0[W]\tr[{\bf b}_4(\bar\Delta)]s^4
+\cdots   \right],
\end{align}
where we have used the formula
\begin{align}
\sum_m \delta(z-\lambda_m) = \int_{\gamma-i\infty}^{\gamma+i\infty}\frac{\df s}{2\pi i} \,\sum_m e^{s(z-\lambda_m)}
=\int_{\gamma-i\infty}^{\gamma+i\infty}\frac{\df s}{2\pi i}\,e^{sz}\, \tr \left[e^{-s \bar\Delta} \right],
\end{align}
with $\gamma$ being real positive finite, and have assumed the eigenvalues of $\bar\Delta$ to be positive,
 i.e., $\lambda_m>0$. The heat kernel expansion has been performed as
\begin{align}
\tr \,\left[e^{-s\bar\Delta}\right] = \frac{1}{(4\pi s)^{2}} \int_x \sqrt{\bar g}
\left[ \tr[{\bf b}_0(\bar\Delta)]  +\tr[{\bf b}_2(\bar\Delta)]s + \tr[{\bf b}_4(\bar\Delta)]s^4+\cdots   \right].
\label{app:eq:heat kernel}
\end{align}
The inverse Laplace transformation yields
\begin{align}
\int_{\gamma-i\infty}^{\gamma+i\infty}\frac{\df s}{2\pi i} \,e^{sz} s^{-n}
=\begin{cases}
\displaystyle \frac{z^{n-1}}{\Gamma(n)} & \text{for $n>0$},\\[2ex]
\delta(z) & \text{for $n=0$},\\[2ex]
\displaystyle \frac{\delta^{|n|}}{\delta z^{|n|}}\delta(z) & \text{for $n<0$},
\end{cases}
\label{eqapp: Laplace trans}
\end{align}
where $\Gamma(n)$ is the Gamma function.
Then, the threshold functions are defined to be
\begin{align}
&Q_n[W] \equiv \frac{1}{\Gamma(n)}\int_0^\infty \df z \,z^{n-1} W[z]\qquad (n > 0),\\
&Q_0[W] \equiv \int_0^\infty \df z \,W[z]\delta(z) = W[0],\\
&Q_{-|n|}[W] \equiv (-1)^{-n}\frac{\df^{|n|} W}{\df z^{|n|}}\qquad (n < 0).
\end{align}

Here, the heat kernel coefficients of the first three terms for $\bar\Delta = -\bar\square - {\bf U}$ are found
to be~\cite{Gilkey:1995mj}
\begin{align}
&{\bf b}_0(\bar\Delta) = {\bf 1},\\
&{\bf b}_2(\bar\Delta) = \frac{\bar R}{6} {\bf 1} - {\bf U},\\
&{\bf b}_4(\bar\Delta) = 
 \left( \frac{1}{180}\bar R_{\mu\nu\rho\sigma}^2 -\frac{1}{180}\bar R_{\mu\nu}^2 + \frac{1}{72}\bar R^2
 -\frac{1}{30}\bar{\square} \bar R \right){\bf 1} \nn
&\phantom{{\bf b}_4(\bar\Delta)}\qquad
 +\frac{1}{2}{\bf U}^2 + \frac{1}{6}\bar{\square} {\bf U} + \frac{1}{12}\Omega_{\mu\nu}\Omega^{\mu\nu}
- \frac{\bar R}{6}{\bf U},
\end{align}
with $\Omega_{\mu\nu}=[\bar\nabla_\mu, \bar\nabla_\nu]$.

\subsection{Scalar-tensor sector}
The identity matrix acting on the field space is given by
 $(h_{\mu\nu}, \varphi)$
\begin{align}
 {\bf 1} = \pmat{
\d^{\a\b}_{\mu\nu} &  0\\
0 & 1
 },
\end{align}
whose trace is
\begin{align}
\tr[{\bf 1}]
=\frac{1}{2}(\d^\a_\mu \d^\b_\nu+\d^\a_{\nu} \d^\b_\mu) \delta_\a^\mu  \delta_\d^\nu
+ 1
=10+1=11.
\end{align}
In the present work, we need to evaluate the heat kernel coefficients for the differential operator $\mathcal D
=-\bar\square+ {\mathbb W}$.
Here ${\mathbb W}$ is given in \Cref{eq:H kernel}. Thus, we evaluate the heat kernel coefficients: 
\begin{align}
\label{eqApp:heat1}
&\tr[{\bf b}_0(\mathcal D)]  = 11,\\
\label{eqApp:heat2}
&\tr[{\bf b}_2(\mathcal D)] = \frac{11}{6}\bar R - \tr[{\mathbb W}],\\
&\tr[{\bf b}_4(\mathcal D)] = 
 \left( \frac{11}{180}\bar R_{\mu\nu\rho\sigma}\bar R^{\mu\nu\rho\sigma} -\frac{11}{180}\bar R_{\mu\nu}\bar R^{\mu\nu} + \frac{11}{72}\bar R^2 -\frac{11}{30}\bar \square \bar R \right) \nn
&\phantom{{\bf b}_4(\mathcal D)}\qquad\quad
 +\frac{1}{2}\tr[{\mathbb W}^2] 
 + \frac{1}{6}\tr[\bar\nabla^2 {\mathbb W}] + \frac{1}{12}\tr[\Omega_{\mu\nu}\Omega^{\mu\nu}]- \frac{\bar R}{6}\tr[{\mathbb W}].
 \label{eq:D4 hkc}
\end{align}
For computing the trace, we use the \textit{Mathematica} packages ``xAct"~\cite{Martin-Garcia:2007bqa,Martin-Garcia:2008ysv,Brizuela:2008ra,Nutma:2013zea}. 
First, we obtain
\begin{align}
\tr[{\mathbb W}]
&
= -m^2 - 10\kappa\varrho  -5 m^2 \kappa \bar\phi^2 +6 \bar R
- \kappa \p_\mu \bar\phi \p^\mu \bar\phi ,\\[2ex]
\tr[{\mathbb W}^2]
&=m^4 +10 \kappa^2 \varrho^2
+m^4\kappa\bar\phi^2
+ 10m^2\kappa^2\varrho\bar\phi^2
-6m^2\kappa \bar\phi^2\bar R
+ \frac{5}{2}m^4\kappa^2\bar\phi^4\nn
&\quad
-12\kappa\varrho \bar R
+3\bar R_{\mu\nu\rho\sigma}\bar R^{\mu\nu\rho\sigma}
+2\bar R_{\mu\nu} \bar R^{\mu\nu}
+ 3 \bar R^2
\nn
&\quad
 +2\kappa (\bar\square\bar\phi)^2
 -4\kappa (\bar\nabla_\mu \bar\nabla_\nu \bar\phi)(\bar\nabla^\mu \bar\nabla^\nu \bar\phi)
\nn
&\quad
-2 \kappa  (\p_\mu\bar\phi )(\p^\mu\bar\phi ) \bar R
+ 8 \kappa \bar R^{\mu\nu} \p_\mu \bar\phi \p_\nu \bar\phi
+ \frac{11}{2}\kappa^2 (\p_\mu\bar\phi \p^\mu\bar\phi)^2 ,\\[2ex]
\tr[\bar \nabla^2{\mathbb W}]
&=  6\, \bar\square\bar R -\kappa \bar\square (\p_\mu \bar\phi \p^\mu \bar\phi).
\end{align}
Next, we evaluate $\tr[\Omega_{\mu\nu}\Omega^{\mu\nu}]$.
Note that in the $(h_{\mu\nu}, \varphi)$-basis, we have
\begin{align}
\Omega_{\mu\nu} = [\bar\nabla_\mu, \bar\nabla_\nu]
=\pmat{
 [\bar\nabla_\mu, \bar\nabla_\nu]_{\alpha\beta \rho\sigma} &  [\bar\nabla_\mu, \bar\nabla_\nu]_{\alpha\beta}\\
  [\bar\nabla_\mu, \bar\nabla_\nu]_{\rho\sigma} &  [\bar\nabla_\mu, \bar\nabla_\nu]_{\cdot\cdot}
}
=\pmat{
(\Omega_{\mu\nu})_{\alpha\beta\rho\sigma} & 0\\
 0 & 0
},
\end{align}
where 
\begin{align}
\tr[\Omega_{\mu\nu}\Omega^{\mu\nu}]
&= \tr\{ [\bar\nabla_\mu, \bar\nabla_\nu][\bar\nabla^\mu, \bar\nabla^\nu] \}\nn
&=\left(
-\bar R_{\mu\nu\gamma\alpha}\bar R^{\mu\nu\gamma}{}_\rho \bar g_{\beta\sigma}
-\bar R_{\mu\nu\gamma\beta}\bar R^{\mu\nu\gamma}{}_\sigma \bar g_{\alpha\rho} + 2\bar R_{\mu\nu\alpha\beta}\bar R^{\mu\nu}{}_{\rho\sigma}
\right)\bar g^{\a\rho}  \bar g^{\b\sigma}\nn
&= -6 \bar R_{\mu\nu\rho\sigma}\bar R^{\mu\nu\rho\sigma}.
\end{align}
Now, by use of the topological Gauss-Bonnet term in four-dimensional spacetimes~\labelcref{eq:Gauss-Bonnet term}, $\bar R_{\mu\nu\rho\sigma}\bar R^{\mu\nu\rho\sigma}$ can be expressed by $\bar R_{\mu\nu}\bar R^{\mu\nu}$ and $\bar R^2$ as
\begin{align}
\bar R_{\mu\nu\rho\sigma}\bar R^{\mu\nu\rho\sigma} = 4\bar R_{\mu\nu}\bar R^{\mu\nu} - \bar R^2
=4\bar S_{\mu\nu}\bar S^{\mu\nu},
\end{align}
up to total derivative term, where we have introduced the traceless tensor $\bar S_{\mu\nu}=\bar R_{\mu\nu}-(\bar R/4)\bar g_{\mu\nu}$.
Using the integration by parts, we have 
\begin{align}
\int_x\sqrt{\bar g}(\bar\nabla_\mu\bar\nabla_\nu \bar\phi)(\bar\nabla^\mu\bar\nabla^\nu \bar\phi)
&=\int_x\sqrt{\bar g} \left[(\bar\square \bar\phi)^2-\br^{\mu\nu}\pa_\mu\bp\, \pa_\nu \bp \right]\nn
&=\int_x\sqrt{\bar g} \left[(\bar\square \bar\phi)^2-\bar S^{\mu\nu}\pa_\mu\bp\, \pa_\nu \bp - \frac{\bar R}{4}(\p_\mu \bar\phi)^2 \right].
\end{align}
Then \Cref{eq:D4 hkc} becomes
\begin{align}
\tr[{\bf b}_4(\mathcal D)]
&= \frac{m^4}{2} + 5\kappa^2 \varrho^2 
+ \frac{1}{2}m^4\kappa\bar\phi^2
+ 5m^2\kappa^2\varrho \bar\phi^2
+ \frac{5}{4}m^4\kappa^2 \bar\phi^4\nn
&\quad
+\frac{m^2}{6}\bar R -\frac{13}{3}\kappa\varrho \bar R
+ \frac{311}{60}\bar S_{\mu\nu}\bar S^{\mu\nu}  + \frac{71}{80}\bar R^2 
\nn
&\quad
-\frac{13}{6}m^2\kappa \bar\phi^2\bar R
+ \frac{2}{3}\kappa (\p_\mu \bar\phi \p^\mu \bar\phi)\bar R
+ 6 \kappa\bar S^{\mu\nu} \p_\mu \bar\phi \p_\nu \bar\phi
+ \frac{11}{4}\kappa^2 (\p_\mu\bar\phi \p^\mu\bar\phi)^2 
\nn
&\quad
-\kappa (\bar\square\bar\phi)^2
+ \frac{19}{30} \bar\square\bar R -\frac{1}{6}\kappa \bar\square (\p_\mu \bar\phi \p^\mu \bar\phi) .
\label{eq:heatkernel D4}
\end{align}

To summarize, the heat kernel coefficients read
\begin{align}
\tr[{\bf b}_2(\mathcal D)]
&= 10\kappa\varrho +5 m^2\kappa \bar \phi^2 -\frac{25}{6} \bar R
+ \kappa \p_\mu \bar\phi \p^\mu \bar\phi,\\
\tr[{\bf b}_4(\mathcal D)]
&= \frac{m^4}{2} + 5\kappa^2 \varrho^2 
+ \frac{1}{2}m^4\kappa\bar\phi^2
+ 5m^2\kappa^2\varrho \bar\phi^2
+ \frac{5}{4}m^4\kappa^2 \bar\phi^4\nn
&\quad
+\frac{m^2}{6}\bar R -\frac{13}{3}\kappa\varrho \bar R
+ \frac{311}{60}\bar S_{\mu\nu}\bar S^{\mu\nu}  + \frac{71}{80}\bar R^2 
\nn
&\quad
-\frac{13}{6}m^2\kappa \bar\phi^2\bar R
+ \frac{2}{3}\kappa (\p_\mu \bar\phi \p^\mu \bar\phi) \bar R
+ 6\kappa \bar S^{\mu\nu} \p_\mu \bar\phi \p_\nu \bar\phi
+ \frac{11}{4}\kappa^2 (\p_\mu\bar\phi \p^\mu\bar\phi)^2 
\nn
&\quad
-\kappa (\bar\square\bar\phi)^2
+ \frac{19}{30} \bar\square\bar R -\frac{1}{6}\kappa \bar\square (\p_\mu \bar\phi \p^\mu \bar\phi) .
\label{eq:heatkernel D4 2}
\end{align}

\subsection{Ghost sector}

Next, we evaluate the heat kernel coefficients for the derivative operator acting on the (spin-1) ghost field
\begin{align}
\Pi^{\mu}{}_\nu&= -\delta^\mu_\nu\bar\square - \kappa \pa^\mu\bp \pa_\nu \bp
-\br^\mu{}_\nu
\equiv  -\delta^\mu_\nu\bar\square + ({\mathbb W}_\text{gh})^\mu{}_\nu\,,
\label{app:eq:ghost diff op}
\end{align}
The identity matrix acting on the spin-1 ghost fields is
\begin{align}
{\bf 1} = \delta^{\mu}_\nu,
\end{align}
and then $\tr[{\bf 1}] = \delta^\mu_\nu \delta^\nu_\mu = 4$.
For \Cref{app:eq:ghost diff op}, the heat kernel coefficients are
\begin{align}
\label{eqapp:ghost heat1}
&\tr[{\bf b}_0(\Pi)]  = 4,\\
\label{eqapp:ghost heat2}
&\tr[{\bf b}_2(\Pi)] = \frac{2}{3}\bar R - \tr[{\mathbb W}_\text{gh}],\\
&\tr[{\bf b}_4(\Pi)] = 
 \left( \frac{4}{180}\bar R_{\mu\nu\rho\sigma}\bar R^{\mu\nu\rho\sigma} -\frac{4}{180}\bar R_{\mu\nu}\bar R^{\mu\nu} + \frac{4}{72}\bar R^2 -\frac{4}{30}\bar\square  \bar R \right) \nn
&\phantom{{\bf b}_4(\mathcal D)}\qquad\quad
 +\frac{1}{2}\tr[{\mathbb W}_\text{gh}^2] 
 + \frac{1}{6}\tr[\bar\square  {\mathbb W}_\text{gh}] + \frac{1}{12}\tr[\Omega^\text{gh}_{\mu\nu}\Omega^\text{gh}{}^{\mu\nu}]- \frac{\bar R}{6}\tr[{\mathbb W}_\text{gh}].
 \label{eqapp:ghost heat3}
\end{align}
Here we compute
\begin{align}
\tr[{\mathbb W}_\text{gh}]
&=  - \kappa \pa^\mu\bp \pa_\mu \bp -\br,\\[1ex]
\tr[{\mathbb W}_\text{gh}^2]&= 2\kappa\bar R_{\alpha\beta}\p^\alpha\bar\phi \p^\beta\bar\phi
+\kappa^2(\p_\alpha \bar\phi \p^\alpha \bar\phi)^2
+\bar R_{\alpha\beta}\bar R^{\alpha\beta},\\[1ex]
\tr[\bar\square{\mathbb W}_\text{gh}]
&=  - \kappa \bar\square(\pa^\mu\bp \pa_\mu \bp)  -\bar\square\br,\\[1ex]
\tr[\Omega^\text{gh}_{\mu\nu}\Omega^\text{gh}{}^{\mu\nu}]
&= \tr\{ [\bar\nabla_\mu, \bar\nabla_\nu][\bar\nabla^\mu, \bar\nabla^\nu] \}\nn
&= -\bar R_{\mu\nu\alpha\rho}\bar R^{\mu\nu\alpha\sigma} \delta^{\rho}_\sigma  \nn
&= - \bar R_{\mu\nu\rho\sigma}\bar R^{\mu\nu\rho\sigma}.
\end{align}
Thus, we obtain
\begin{align}
&\tr[{\bf b}_2(\Pi)] = \kappa (\pa^\mu\bp \pa_\mu \bp) +\frac{5}{3}\bar R ,\\
&\tr[{\bf b}_4(\Pi)] = \frac{41}{120}\bar R^2 + \frac{7}{30}\bar S^{\mu\nu}\bar S_{\mu\nu}
+\frac{5}{12}\kappa \bar R(\p_\mu \bar\phi \p^\mu\bar\phi) + \kappa \bar S^{\mu\nu}(\p_\mu \bar\phi \p^\nu\bar\phi) \nn
&\phantom{\tr[{\bf b}_4(\Pi)]=}
+ \frac{1}{2}\kappa^2 (\p_\mu\bar\phi \p^\mu \bar\phi)^2
-\frac{3}{10}\bar\square \bar R.
\end{align}

\section{Equations of motion}
\label{eom}

The background-field equations from the action~\labelcref{eq:starting effect action} are derived by
the first variations such that
\begin{align}
\label{eq1}
&0=\frac{1}{\sqrt{\bar g}}\frac{\delta S}{\delta \bar\phi}
= \bar\square \, 
\bar\phi + m^2\bar\phi \,,\\[1ex]
&0=\frac{1}{\sqrt{\bar g}}\frac{\delta S}{\delta \bar g^{\mu\nu}}
=-\frac{1}{2}\varrho \, \bar g_{\mu\nu}-\frac{1}{16\pi G_N}\left(\bar R_{\mu\nu}
-\frac12 \bar g_{\mu\nu} \bar R\right)-\frac12 \p_\mu  \bar\phi \p_\nu \bar\phi \nn
& \phantom{0=\frac{1}{\sqrt{\bar g}}\frac{\delta S}{\delta \bar g^{\mu\nu}}=} \quad
+\frac14 \bar g_{\mu\nu} \left(\bar g^{\rho\lambda}
\p_\rho \bar\phi \p_\lambda \bar\phi \right) -\frac14 m^2 \bp^2 \bg_{\mu\nu}.
\label{eq2}
\end{align}
From \Cref{eq1}, we have 
\begin{align}
\left(\bar\square + m^2 \right)\bar\phi =0.
\end{align}
Taking the trace on \Cref{eq2}, we find
\begin{align}
\bar g^{\rho\lambda}
\p_\rho \bar\phi \p_\lambda \bar\phi
= \p_\mu \bar\phi \p^\mu\bar\phi
= 4\varrho - \frac{1}{8\pi G_N}\bar R +2m^2\bp^2,
\end{align}
and plugging this into \Cref{eq2} yields
\begin{align}
\frac12 \varrho \bg_{\mu\nu}-\frac{1}{16\pi G_N} \left(\bar S_{\mu\nu}+\frac14 \br \bg_{\mu\nu}\right) -\frac12 \pa_\mu \bp \pa_\nu \bp +\frac14 m^2 \bp^2 \bg_{\mu\nu}=0.
\end{align}

\bibliographystyle{JHEP} 
\bibliography{refs}
\end{document}